\documentclass[traditabstract]{aa} 
\usepackage{graphicx}
\usepackage{txfonts}
\def\ms{\hbox{\,m\,s$^{-1}$}}         
\def\m2s2{\hbox{\,m$^{2}$\,s$^{-2}$}} 
\def\kms{\hbox{\,km\,s$^{-1}$}}       
\def\vsini{\hbox{$v$\,sin\,$i$}}      

\def\Mjup{\hbox{$\mathrm{M}_{\rm Jup}$}}
\def\Rjup{\hbox{$\mathrm{R}_{\rm Jup}$}}

\def \1s{$1\,\sigma$}

\def \t0{T$_0$}

\begin{document}
   \title{WASP-21b: a hot-Saturn exoplanet transiting a thick disc star\thanks{Based on observations 
   made with the SuperWASP-North camera hosted by the Isaac Newton Group on La Palma, the FIES
   spectrograph on the Nordic Optical Telescope, the CORALIE spectrograph on the 1.2-m Euler Swiss telescope on
   La Silla Observatory, the SOPHIE spectrograph on the 1.93-m telescope on Haute Provence Observatory
   and the HARPS spectrograph on the 3.6-m ESO telescope at La Silla Observatory under  
   programs 081.C-0388, 082.C-0040, 084.C-0185. Tables of photometric data are only available 
   in electronic form at the CDS via anonymous ftp to cdsarc.u-strasbg.fr (130.79.128.5) 
   or via http://cdsweb.u-strasbg.fr/cgi-bin/qcat?J/A+A/.}}


\author{
F. Bouchy, \inst{1,2}
\and L. Hebb, \inst{3,5}
\and I. Skillen, \inst{4}
\and A. Collier Cameron, \inst{5}
\and B. Smalley, \inst{6}
\and S. Udry, \inst{7}
\and D.R. Anderson, \inst{6}
\and I. Boisse, \inst{1}
\and B. Enoch, \inst{5}
\and C.A. Haswell, \inst{8}
\and G. H\'ebrard, \inst{1}
\and C. Hellier, \inst{6}
\and Y. Joshi, \inst{9}
\and S.R. Kane, \inst{10}
\and P.F.L. Maxted, \inst{6}
\and M. Mayor, \inst{7}
\and C. Moutou, \inst{11} 
\and F. Pepe, \inst{7}
\and D. Pollacco, \inst{9}
\and D. Queloz, \inst{7}
\and D. S\'egransan, \inst{7}
\and E.K. Simpson, \inst{9}
\and A.M.S. Smith,  \inst{6}
\and H.C. Stempels, \inst{12}
\and R. Street, \inst{13}
\and A.H.M.J. Triaud, \inst{7}
\and R.G. West, \inst{14}
\and P.J. Wheatley. \inst{15}
}

\institute{
Institut d'Astrophysique de Paris, UMR7095 CNRS, Universit\'e Pierre \& Marie Curie, 
98bis Bd Arago, 75014 Paris, France
\and
Observatoire de Haute-Provence, CNRS/OAMP, 04870 St Michel l'Observatoire, France
\and 
Department of Physics and Astronomy, Vanderbilt University, Nashville, TN 37235, USA
\and
Isaac Newton Group of Telescopes, Apartado de Correos 321, E-38700 Santa 
Cruz de la Palma, Tenerife, Spain 
\and  
SUPA, School of Physics and Astronomy, University of St Andrews, North Haugh, St Andrews, 
Fife KY16 9SS, UK
\and
Astrophysics Group, Keele University, Staffordshire, ST5 5BG, UK
\and
Observatoire de Gen\`eve, Universit\'e de Gen\`eve, 51 Ch. des Maillettes, 1290 Sauverny, 
Switzerland
\and
Department of Physics and Astronomy, The Open University, Milton Keynes, MK7 6AA, UK
\and
Astrophysics Research Centre, School of Mathematics and Physics, Queen's 
University, University Road, Belfast, BT7 1NN, UK 
\and
NASA Exoplanet Science Institute, Caltech, MS 100-22, 770 South Wilson Avenue, Pasadena, CA 91125, USA
\and 
Laboratoire d'Astrophysique de Marseille, 38 rue Fr\'ed\'eric Joliot-Curie, 
13388 Marseille cedex 13, France
\and 
Department of Physics and Astronomy, Uppsala University, Box 516, 75120 Uppsala, Sweden
\and
Las Cumbres Observatory, 6740 Cortona Drive Suite 102, Goleta, CA 93117, USA
\and
Department of Physics and Astronomy, University of Leicester, Leicester, LE1 7RH 
\and
Department of Physics, University of Warwick, Coventry CV4 7AL, UK
}

\date{Received ; accepted }

 
\abstract
{We report the discovery of WASP-21b, a new transiting exoplanet discovered by the Wide Angle Search 
for Planets (WASP) Consortium and established and characterized with the FIES, SOPHIE, CORALIE and HARPS 
fiber-fed echelle spectrographs. A 4.3-d period, 1.1\% transit depth and 3.4-h duration are derived 
for WASP-21b using SuperWASP-North and high precision photometric observations at the Liverpool Telescope.
Simultaneous fitting to the photometric and radial velocity data with a Markov Chain Monte Carlo
procedure leads to a planet in the mass regime of Saturn. With a radius of 1.07 {\Rjup} and mass of 
0.30 \Mjup, WASP-21b has a density close to 0.24 $\rho_{Jup}$ corresponding to the distribution peak 
at low density of transiting gaseous giant planets. With a host star metallicity $[Fe/H]$ of -0.46, 
WASP-21b strengthens the correlation between planetary density and host star metallicity 
for the five known Saturn-like transiting planets. Furthermore there are clear indications that 
WASP-21b is the first transiting planet belonging to the thick disc. }

\keywords{planetary systems -- techniques: photometric -- Techniques : radial velocities}

   \maketitle
%

\section{Introduction}

Observations of planets that transit their host star represent currently the best 
opportunity to test models of exoplanet structure and evolution. These last ten years, 
the photometric surveys have led to an increasing list of transiting planets. 
More than seventy transiting planetary systems have been identified from Super-Earth 
to Jupiter-like planets. The WASP project operates two identical instruments, at La 
Palma in the Northern hemisphere, and at Sutherland in South Africa in the Southern 
hemisphere and led recently to the detection of about 30\% of known transiting 
planets. Each WASP telescope has a field of view of just under 500 square degrees. 
The WASP survey is sensitive to planetary transit signatures in the light curves of 
stars in the magnitude range V $\sim$ 9-13. A detailed description of the telescope hardware, 
observing strategy and pipeline data analysis is given in Pollacco et al. (\cite{pollacco06}). 

Here we present the WASP photometry of SWASPJ230958.23+182346.0, high-precision photometric 
follow-up observations with the RISE instrument on the Liverpool Telescope and high-precision radial
velocity (RV) observations with the FIES (2.6-m NOT), SOPHIE (1.93-m OHP), CORALIE (1.2-m Euler) 
and HARPS (3.6-m ESO) fiber-fed echelle spectrographs. These observations lead to the discovery of 
WASP-21b, a hot-Saturn transiting exoplanet.  

\section{Observations}

\subsection{SuperWASP observation}

\begin{figure}
\centering
\includegraphics[width=8.5cm]{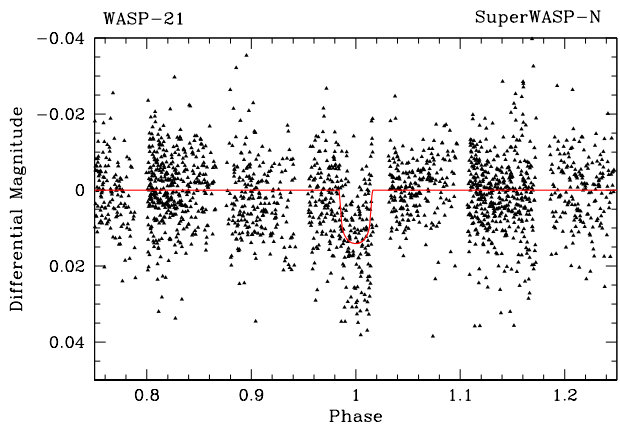}
   \caption{Un-binned SuperWasp folded light curve for WASP-21 with the 
   ephemeris P=4.32 days and T0=2454743.0419.}
    \label{wasp_lc}
\end{figure}

SuperWASP-North is a multi-camera telescope system located 
in La Palma and consisting of 8 Canon 200-mm f/1.8 lenses each coupled 
to e2v 2048$\times$2048 pixel back illuminated CCDs (Pollacco et al. \cite{pollacco06}). 
This combination of lens and camera yields a field-of-view of 
7$^o$.8 $\times$ 7$^o$.8 with an angular size of 13.7 arcsec per pixel.
The target SWASPJ230958.23+182346.0 was monitored from 2006 July 22
to November 27, totaling 3814 photometric measurements (available in electronic form at CDS). 
The pipeline-processed data were detrended and searched for transits with the methods 
described in Collier Cameron et al. (\cite{cameron06}), yielding a detection of a periodic 
transit-like signature with a period of 4.32 days. 
Figure \ref{wasp_lc} shows the un-binned SuperWASP phase-folded light curve 
of WASP-21 with the ephemeris P=4.32 days and T0=2454743.0419.\\

\subsection{Higher precision photometric follow-up} 

\begin{figure}
\centering
\includegraphics[width=8.5cm]{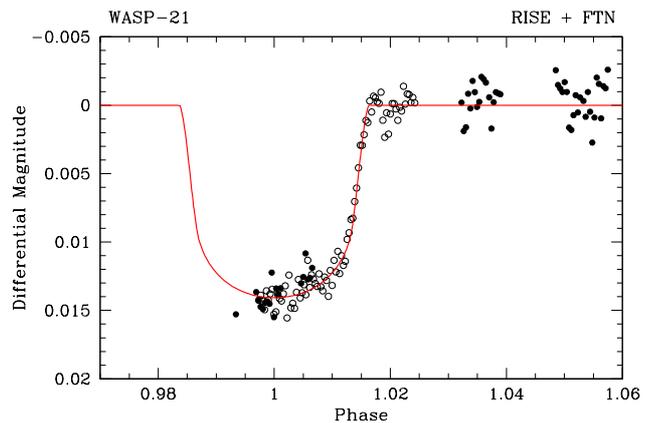}
   \caption{Phase-folded light curve from the RISE (open circle) and FTN (black circle) 
   photometry for WASP-21 superposed with the MCMC solutions. The RISE photometry was binned 
   2-minutes in time.}
     \label{rise_ftn_lc}
\end{figure}

On-off photometric observations were made with the 2-m Faulkes Telescope North (FTN)
at Hawaii on 2th September 2008 with the Pan-STARRS-z broad-band filter. These 
observations allowed the transit signal to be confirmed. Follow-up photometry was 
performed with the fast read-out camera RISE located on the 2.0-m Liverpool 
Telescope at La Palma. Although the night was non-photometric and the sky was 
variable, a partial transit was observed on 7th October 2008 in agreement with 
the SuperWasp and FTN photometry. The phase-folded light curve of FTN and RISE 
(available in electronic form at CDS) are shown in Fig.~\ref{rise_ftn_lc}. 

On October 2009, out-of-transit high-angular resolution images were provided on the 
4.2-m WHT with the Adaptive Optic NAOMI and the IR imager INGRID. 
Ks-band images have a full width at half maximum (FWHM) of 0.18 arcsec with circular 
contours, excluding 
evidence of a visible companion on that spatial scale. The limit in magnitude 
is estimated to be Ks$\sim$13.2 at 0.36 arcsec and to be Ks$\sim$16.8 away from 
the tail of the point spread function (PSF).

\subsection{Doppler follow-up}

Three spectra were first obtained at the end of December 2007 with the FIES spectrograph
at the 2.6-m Nordic Optical Telescope (NOT - La Palma). The combined spectrum indicated a dwarf star and the 
radial velocities showed no variations greater than 100 {\ms}, excluding a low-mass star 
companion. Radial velocity (RV) follow-up of WASP-21 was conducted with the echelle spectrograph 
SOPHIE at the 1.93-m telescope of Observatoire de Haute Provence (Perruchot et al. \cite{perruchot08}, 
Bouchy et al. \cite{bouchy09}), CORALIE at the 1.2-m Swiss Euler telescope of La Silla (Baranne et al., 
\cite{baranne96}, Queloz et al. \cite{queloz00}, Pepe et al. \cite{pepe02}), 
and HARPS at the 3.6-m telescope of ESO La Silla (Mayor et al. \cite{mayor03}).    
A total of eleven SOPHIE measurements were made in July and August 2008 with  
about the same signal-to-noise ratio S/N=30 in order 
to minimize the charge transfer inefficiency (CTI) effect (Bouchy et al. \cite{bouchy09}) 
although it is now corrected by the data reduction software (DRS). Some measurements were 
made with moon light but 
without significant effect on the RV measurements considering that the RV of the 
target (-89.4 \kms) was always far from the moon RV. The RV uncertainties 
were computed assuming photon noise plus 15 {\ms} quadratically added  
in order to take into account the systematics of the high efficiency mode 
like the guiding and centering errors (Boisse et al. \cite{boisse10}) and wavelength calibration 
uncertainty.  
Five CORALIE measurements were made simultaneously with those of SOPHIE and eight additional 
ones were made one year later in June and September 2009. 
One CORALIE point was made close to the transit epoch (BJD=2454686.83) and may be affected by 
the Rossiter-McLaughlin effect. We did not use it to determine the Keplerian 
solution. However, considering the small {\vsini} ($\le$ 2 \kms) of the star, the 
expected amplitude of the RM effect should be lower than 20 {\ms}, which makes 
it difficult to detect with CORALIE or SOPHIE. 
Nine HARPS measurements were made in September and October 2008 to definitively secure 
the candidate considering the small RV amplitude observed by SOPHIE and CORALIE, and also 
to exclude blended eclipsing binaries scenario and to provide a high S/N spectrum for 
stellar parameters analysis (see Sect.~\ref{secstellar}). Six HARPS measurements 
were added in October 2009 to check for eccentricity and for additional companion 
in the system. 

\begin{figure}
\centering
\includegraphics[width=8.5cm]{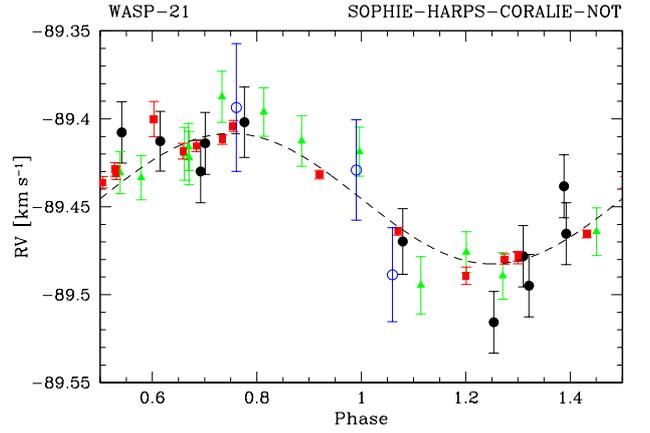}
\caption{Phase-folded radial velocities of WASP-21 obtained with SOPHIE (black circles), 
HARPS (red squares), CORALIE (green triangles) and FIES (blue open circles) }
\label{figrv}
\end{figure}

The phase-folded RVs in Fig.~\ref{figrv} shows the SOPHIE (black circles), HARPS (red squares), 
CORALIE (green triangles) and FIES (blue open circles) points. All are phase matched pretty well 
with the photometry. The best Keplerian fit was obtained with 
an RV offset from the HARPS data of +11, -20.5 and +15 {\ms} for the 
CORALIE, SOPHIE and FIES data respectively. The semi-amplitude of the 
RV curve is $K$=37.2$\pm$1.1 \ms. We adjust the Keplerian orbit fixing the eccentricity 
to zero. There is no obvious evidence for eccentricity, although it is difficult to fit 
for the eccentricity with a small amount of data points from four different instruments 
because we already have three other free parameters (the zero-point offsets of the RV data sets).

No significant RV drift was detected with HARPS during one year excluding additional 
Jupiter-like companion with a period of less than one year. 

\begin{table}[h]
  \centering 
  \caption{Radial velocities of WASP-21}
  \label{table_rv}
\begin{tabular}{ccc}
\hline
\hline
BJD & RV & $\pm$$1\,\sigma$  \\
-2\,400\,000 & (km\,s$^{-1}$) & (km\,s$^{-1}$)  \\
\hline
\multicolumn{3}{c}{FIES - 2.6m NOT} \\
\hline
54462.336 &	-89.504 &	0.027	\\
54465.371 &	-89.409 &	0.036	\\
54466.364 &	-89.444 &	0.029	\\
\hline 
\multicolumn{3}{c}{SOPHIE - 1.93m OHP} \\
\hline
54659.5872   &   -89.409   &    0.018	 \\  
54662.5947   &   -89.418   &    0.018	 \\  
54663.5749   &   -89.392   &    0.017	 \\  
54665.5826   &   -89.449   &    0.019	 \\  
54666.5773   &   -89.458   &    0.017	 \\  
54667.5790   &   -89.387   &    0.017	 \\  
54668.5925   &   -89.381   &    0.020	 \\  
54679.5940   &   -89.474   &    0.018	 \\  
54683.6242   &   -89.495   &    0.018	 \\  
54685.5591   &   -89.393   &    0.018	 \\  
54688.5479   &   -89.445   &    0.018	 \\  
\hline 
\multicolumn{3}{c}{CORALIE - 1.2m Euler} \\
\hline
54624.9095 &	-89.427 &	0.013		\\
54661.7883 &	-89.487 &	0.012		\\
54662.8643 &	-89.475 &	0.014		\\
54684.8541 &	-89.442 &	0.012		\\
54686.8333 &	-89.430 &	0.014		\\
55001.9020 &	-89.424 &	0.014		\\
55009.8829 &	-89.398 &	0.015		\\
55013.8951 &	-89.431 &	0.015		\\
55093.6551 &	-89.506 &	0.016		\\
55095.6656 &	-89.444 &	0.013		\\
55096.6795 &	-89.407 &	0.014		\\
55098.6554 &	-89.500 &	0.013		\\
55130.6417 &	-89.433 &	0.015		\\
\hline
\multicolumn{3}{c}{HARPS - 3.6m ESO} \\
\hline
54706.7464 &	-89.4001 &	0.0100	\\
54709.7655 &	-89.4792 &	0.0033	\\
54710.7472 &	-89.4287 &	0.0038	\\
54749.6611 &	-89.4309 &	0.0036	\\
54754.6434 &	-89.4154 &	0.0031	\\
54755.6635 &	-89.4317 & 	0.0025	\\
54760.6340 &	-89.4639 &	0.0022	\\
54761.6280 &	-89.4782 & 	0.0027	\\
54763.5963 &	-89.4040 &	0.0031	\\
55111.6377 &	-89.4797 &	0.0029	\\
55112.6342 &	-89.4361 &	0.0032	\\
55113.6229 &	-89.4118 &	0.0027	\\
55115.6429 &	-89.4892 &	0.0050	\\
55116.6385 &	-89.4655 & 	0.0023	\\
55117.6204 &	-89.4184 &	0.0045	\\
\hline
\end{tabular}
\end{table}

\begin{figure}
\centering
\includegraphics[width=8.5cm]{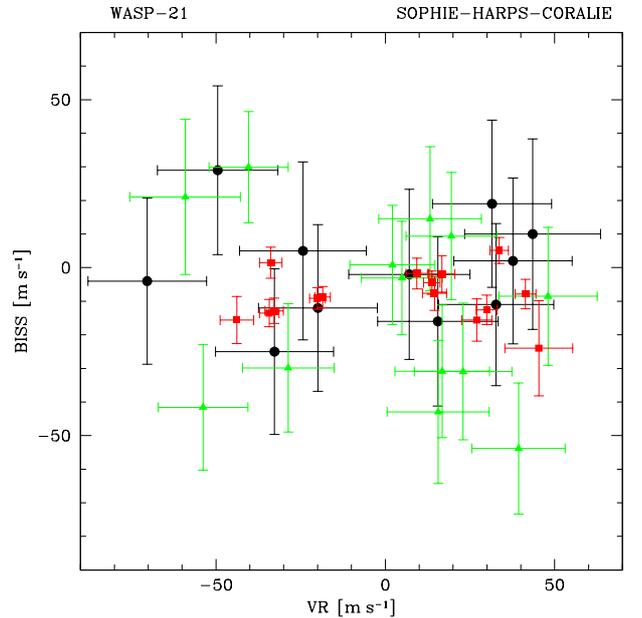}
   \caption{Bisector span versus radial velocities of WASP-21 obtained with SOPHIE (black circles), 
HARPS (red squares) and CORALIE (green triangles) without significant correlation and
   invalidating scenarios of blended eclipsing binaries.}
\label{figbis}
\end{figure}

Figure 4 shows the bisector span versus RV. The SOPHIE and CORALIE bisector
spans have too large uncertainties to exclude a possible correlation with radial
velocities. The HARPS bisector spans definitively show no correlation, and
scenarios of blended eclipsing binaries are definitively invalided.

\section{Results and parameters}

\subsection{Stellar parameters}
\label{secstellar}

A detailed spectroscopic analysis of the stellar atmospheric properties was made 
with the FIES and HARPS spectra. We merged the HARPS spectra into one high-quality 
spectrum in order to perform a detailed spectroscopic analysis of the stellar 
atmospheric properties. This merged spectrum was continuum-normalized with a 
low-order polynomial to retain the shape of the broadest spectral features. 
The total signal-to-noise ratio of the combined spectrum was about 100 per pixel.

As previously done for our analysis of WASP-1 (Stempels et al. \cite{stempels07}) 
and WASP-3 (Pollacco et al. \cite{pollacco08}), we employed the methodology of Valenti \& Fischer 
(\cite{valenti05}), using the same tools, techniques and model atmosphere grid. We used the 
package Spectroscopy Made Easy (SME) (Valenti \& Piskunov \cite{valenti96}), which combines 
spectral synthesis with multidimensional $\chi^2$ minimization to determine which 
atmospheric parameters best reproduce the observed spectrum of WASP-21 (effective 
temperature $T_{eff}$, surface gravity $\log g$, metallicity $[M/H]$, projected {\vsini}, 
microturbulence $v\_{mic}$ and the macroturbulence $v\_{mac}$). The parameters we 
obtained from this analysis are listed in Table~\ref{stellarparam}.

We interpolated the Girardi evolutionary tracks using a 12 Gyr old star with 
$T_{eff}$=5800 and $[M/H]$=-0.4 and found $Mv=4.97$. A reddening $E(B-V)$ of 0.1 was 
estimated from the equivalent width of interstellar NaD lines with the 
Munari \& Zwitter (\cite{munari97}) relationship. With an observed Vmag of 11.58,  
the distance is estimated to be 242$^{+50}_{-30}$ pc. This value agrees with the distance 
of 235 pc derived through the angular diameter ($\theta$ = 0.042 $\pm$ 0.002 mas) 
estimated from the infrared flux method and the stellar radius from Table~\ref{mcmc}.
It also agrees with the distance of $230~\pm 30$~pc 
based on the measured V band magnitude ($V=11.58\pm0.08$) and the absolute V~magnitude 
of a typical G3V type star (Gray \cite{gray88}).

The properties of WASP-21 suggest it is a member of the thick disc component of the Galaxy.  
With the NOMAD catalogue proper motions (Zacharias et al. \cite{NOMAD}) and the measured 
systemic radial velocity, we calculated its $U$, $V$, and $W$ space motion to be 
-34.6, -66.0, +62.1 {\kms} relative to the Sun, adopting a distance of 242~pc 
based on the measured V band magnitude ($V=11.58\pm0.08$) and the absolute V~magnitude 
of a typical G3V type star (Gray \cite{gray88}). It is clear that WASP-21 is lagging behind the 
Sun (which moves with the thin disc) with a high negative $V$ velocity similar to the 
canonical thick disc value of $\sim$ -45 {\kms} (Morrison et al. \cite{MFF90}).  
WASP-21 is also metal-poor with $[M/H]=-0.4\pm0.1$, again similar to the mean thick disc 
value of $\sim -0.6$~dex.
In addition, the abundance analysis shows enhanced $\alpha$/Fe ratios, which is also typical 
of a thick disc star (see Edvardsson et al. \cite{edvardsson93}). The alpha elements,  Mg
Ca, Ti, and Si have abundances of $\sim$-0.25, approximately twice the value of iron, 
$[Fe/H]$ =-0.46, and the other iron peak elements ($[V/H]$=-0.32, $[Cr/H]$=-0.44, $[Mn/H]$=-0.75, 
$[Co/H]$=-0.39 and $[Ni/H]$=-0.46). 
Furthermore, when we compare the effective temperature ($T_{eff} = 5800\pm 100$~K), metallicity 
and mean stellar density ($\rho/\rho_{\odot}=0.84$) derived from the transit parameters of 
WASP-21 to theoretical stellar evolution models from Girardi et al. (\cite{girardi00}) as described 
in Hebb et al. (\cite{hebb09}), we find the star to be evolved off the zero-age main sequence and 
similar in age to that of the thick disc (age $\sim 12 \pm 5$~Gyr).
In order to quantify the probability that WASP-21 is a member of the thick disc, we compared 
its chemical and kinematic properties to the properties of similar model stars in the 
Besan\c{c}on Galactic model (Robin et al. \cite{besancon}) where the population class of the 
stars are known.  
We selected a set of model main sequence stars in a small volume nearby the position of WASP-21 
for comparison. We generated more than 800,000 model stars in 55 realizations of the simulation 
with spectral types of F5V-K7V, distances of 150-600~pc, and positions within $\pm 5^{\circ}$ 
in each direction of WASP-21 ($l=92$,$b=-38$). The model stars come from all Galactic 
components and are not selected based on their velocities or metallicities. Of the model stars 
with metallicities between -0.5 and -0.3~dex and within $\pm 15$~{\kms}
of the calculated  $U$,$V$, and $W$ values of WASP-21, all have ages $\ge 7$~Gyr (population 
class 7 and 8) which suggests WASP-21 is old. Furthermore, 92\% of the matching model stars 
belong to the thick disc.
In summary, WASP-21 has a high probability of being a thick disc member based
on its kinematic properties, low metallicity, abundance patterns, and likely old age.

\begin{table}
\begin{center}
\caption[]{Stellar parameters for WASP-21 derived from FIES and HARPS spectroscopy.}
\label{stellarparam}
\begin{tabular}{lc}
\hline
\hline
Parameter  & WASP-21  \\
\hline
GSC & 01715-00679 \\
2MASS &  J23095825+1823459  \\
RA (J2000)	&  23:09:58.25 \\
DEC (J2000)	&  +18:23:45.9 \\
V		&   11.58$\pm$0.08   \\
Distance 	&   242$^{+50}_{-30}$ pc \\
$T_{eff}$          &           5800 $\pm$ 100      K\\
$\log g$         &           4.2  $\pm$ 0.1   \\
$[M/H]$         &          -0.4  $\pm$ 0.1   \\
\vsini         &           1.5  $\pm$ 0.6    km/s\\
$v_{rad}$	& 	-89.45 \kms \\
Age        & 12 $\pm$ 5 Gyr  \\
\hline
$\log$ A(Li)     &           2.19 $\pm$ 0.09  \\
$[Na/H]$         &         -0.47 $\pm$ 0.11  \\
$[Mg/H]$         &         -0.28 $\pm$ 0.08  \\
$[Ca/H]$         &         -0.25 $\pm$ 0.13  \\
$[Ti/H]$         &         -0.28 $\pm$ 0.10  \\
$[Mn/H]$         &         -0.75 $\pm$ 0.14  \\
$[Fe/H]$         &         -0.46 $\pm$ 0.11  \\
$[Si/H]$         &         -0.25 $\pm$ 0.12  \\
$[Sc/H]$         &         -0.33 $\pm$ 0.11  \\
$[V/H] $         &         -0.32 $\pm$ 0.10  \\
$[Cr/H]$         &         -0.44 $\pm$ 0.17  \\
$[Co/H]$         &         -0.39 $\pm$ 0.14  \\
$[Ni/H]$         &         -0.46 $\pm$ 0.14  \\
\hline
\end{tabular}
\end{center}
\end{table}


\subsection{Planet parameters}

To determine the planetary and orbital parameters the radial velocity measurements 
were combined with the photometry from WASP, RISE and FTN in a simultaneous fit 
with the Markov Chain Monte Carlo (MCMC) technique. The details of this process 
are described in Pollacco et al. (\cite{pollacco08}). Recent features and improvements 
were included. The linear de-trending of the optical light curves was made 
with respect to phase at each step in the MCMC chain. An initial fit showed that the 
orbital eccentricity was poorly constrained by the available data and nearly consistent 
with zero. We therefore fixed the eccentricity parameter at zero.
We derived the host star mass consistently within the MCMC code by comparing the mean 
stellar density (measured from the shape of the transit) and the observed metallicity 
and effective temperature of the host star to an empirical relation defined by a set 
of detached eclipsing binaries with component masses and radii measured to high 
precision ($\le 3$\%) (Torres et al. \cite{torres10}). With this novel technique, 
we derived a mass for WASP-21 of $M = 1.01 \pm 0.03 M_{\odot}$.  The mass and radius of
WASP-21b, given in Table~\ref{mcmc} with the other best-fit parameters,
are then found to be $M_P$= 0.300$\pm$0.011 {\Mjup} and $R_P$ = 1.07$\pm$0.06 {\Rjup}, respectively.

Finally, we note that the host star mass derived from the theoretical stellar
evolution models of Girardi et al. (\cite{girardi00}) is significantly less than what we find
using the empirical relation, even though we used the same observed stellar density, temperature
and metallicity. Upon further investigation, we found that the
metallicity has a strong effect on the radii and effective temperatures of the theoretical
stars, which is not present in the observed eclipsing binary systems.
Specifically, a $\sim 5800$~K, solar metallicity ($Z=0.019$) theoretical star with an age 
of 5~Gyr has a predicted mass of $1.0 M_{\odot}$, but at $Z=0.008$ ($[M/H]=-0.37$), the 
predicted mass is $\sim 10$\% less ($0.9 M_{\odot}$). However, the four eclipsing binary 
components in Torres et al. (\cite{torres10}) with temperatures within $\pm 100$~K of WASP-21, 
have [Fe/H] measurements ranging 
from -0.1 to +0.24, but masses similar to within 5\%. There is a clear difference between the 
predicted effects of metallicity on the properties of stars and the observed effects in 
eclipsing binaries. Using a stellar mass of $0.9 M_{\odot}$ leads to a planetary mass and radius
for WASP-21b of $M_P$= 0.29$\pm$0.01 {\Mjup} and $R_P$ = 1.03$\pm$0.06 {\Rjup}, respectively,  
$\sim 3$\% less than our adopted value.

\begin{table}
\begin{center}
\caption[]{WASP-21 system parameters and 1$\sigma$ error limits derived
from the MCMC analysis.}
\label{mcmc}
\begin{tabular}{lccl}
\hline
\hline
Parameter & Value  \\
\hline
Transit epoch $T_0$ [HJD] & $ 2454743.0419^{+ 0.0019}_{- 0.0022} $ \\
Orbital period $P$ [days] & $ 4.322482^{+ 0.000019}_{- 0.000024} $ \\
Planet/star area ratio $ (R_p/R_*)^2 $ & $ 0.01082^{+ 0.00037 }_{- 0.00035} $  \\
Transit duration $t_T$ [days] & $ 0.1398^{+ 0.0048}_{- 0.0040} $ \\
Impact parameter $ b $ [$R_*$]  & $ 0.23^{+ 0.12}_{-0.15} $  \\
Orbital inclination $ I $ [degrees] & $ 88.75^{+0.84}_{- 0.70} $ \\
 &    &      &  \\
Stellar reflex velocity $K$ [\ms] & $ 37.2 \pm 1.1 $ \\
Orbital semimajor axis $ a $ [AU] & $ 0.052^{+ 0.00041}_{- 0.00044} $ \\
 &    &      &  \\
Stellar mass $ M_* $ [$M_\odot$] & $ 1.01 \pm 0.03  $  \\
Stellar radius $ R_* $ [$R_\odot$]  & $ 1.06 \pm 0.04 $  \\
Stellar surface gravity $ \log g_* $ [cgs]   & $ 4.39 \pm 0.03 $  \\
Stellar density $ \rho_* $ [$\rho_\odot$] & $ 0.84 \pm 0.09  $ \\
 &    &      &  \\
Planet mass  $ M_p $ [$M_J$]  & $ 0.300 \pm 0.011 $  \\
Planet radius $ R_p $ [$R_J$]  & $ 1.07 \pm 0.06 $  \\
Planet density  $ \rho_p $ [$\rho_J$]  & $ 0.24 \pm 0.05 $  \\
\hline
\end{tabular}
\end{center}
\end{table}

\section{Discussion and conclusion}

\begin{figure}
\centering
\includegraphics[width=8.5cm]{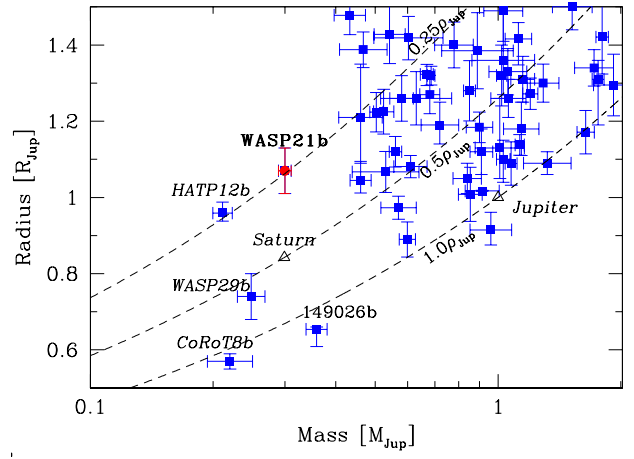}
   \caption{Mass-radius diagram of transiting gaseous giant exoplanets ($M_P \ge$ 0.1 \Mjup).}
      \label{figmr}
\end{figure}

Figure~\ref{figmr} shows the mass-radius diagram of known gas-giant transiting 
exoplanets (with a mass of more than 0.1 \Mjup). WASP-21b is one of the only five transiting 
gas-giant planets with a mass below 0.4 {\Mjup} including HAT-P-12b (Hartman et al. \cite{hartman09}), 
HD149026b (Sato et al. \cite{sato05}), CoRoT-8b (Bord\'e et al. \cite{borde10}) and WASP-29b 
(Hellier et al. \cite{hellier10}). 

\begin{figure}
\centering
\includegraphics[width=8.5cm]{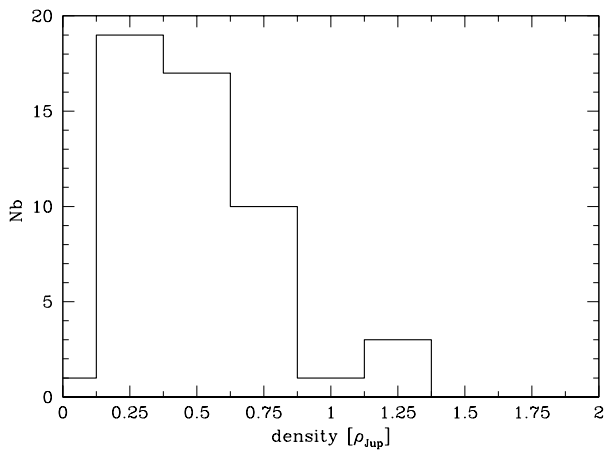}
   \caption{Histogram of the density of transiting hot-Jupiters with a mass in the range 
   0.2-1.5 \Mjup.}
    \label{figdens}
\end{figure}

Figure~\ref{figdens} shows the density histogram of transiting gas-giant planets  
with a mass between 0.2 and 1.5 Jupiter masses (totaling 50 exoplanets). The probability density 
function presents a clear asymmetric distribution peaking at the lowest densities (0.25 $\rho_{Jup}$) 
and decreasing from 0.25 to 1.25 $\rho_{Jup}$. WASP-21b and HAT-P-12b, both with a density of 
$\sim$ 0.24 $\rho_{Jup}$, are bloated planets with a density close to the mode of that 
distribution. This distribution is biased because the transit detection method 
(especially from ground) favours planets 
with large radii, hence with lower density for a given mass. 
If large radii are predominantly the result of irradiation, the selection effects will
favour stars with small $a/R_*$, because that gives a stronger irradiation of the planet.
Small $a/R_*$ also increases the transit detection probability.  

Only four transiting exoplanets with a mass in the range 0.2-1.5 $M_{Jup}$ have 
a density greater than or equal to Jupiter including the two Saturn-like planets CoRoT-8b and 
HD149026b as well as WASP-7b (Hellier et al. \cite{hellier08}) and OGLE-TR-113b (Bouchy et al. 
\cite{bouchy04}), all of them having a host star with metallicity $[Fe/H] \ge 0$. 
The density of the five Saturn-like transiting planets appears to be clearly correlated 
with the metallicity of their host stars ($[Fe/H]$=-0.46, -0.29, +0.11, +0.3 and +0.36 
for WASP-21b and HAT-P-12b, WASP-29b, CoRoT-8b and HD149026b respectively), confirming 
and reinforcing the relation established by Guillot et al. (\cite{guillot06}). 
If low host star metallicity also favours larger planetary radii, it may explain the difference 
observed between the host star metallicity distribution from radial velocity surveys (50\% with
$[Fe/H]\ge0.1$) and transit surveys (40\% with $[Fe/H]\ge0.1$).   

It is clear from previous studies that thick disc stars do form planets, 
which means that planet formation was going on when the Milky Way was
still relatively young (i.e.\ as early as $\sim 12$~Gyr ago)
and that these planetary systems have survived the process (i.e.\ major
merger with a satellite galaxy) that formed the thick disc.
WASP-21b is the first transiting planet with a measured radius and definitive mass 
to orbit a member of this population.

Reid et al. (\cite{reid07}) identify five previously known exoplanet
host stars, which are likely thick disc members based on their subsolar
metallicities, relatively large space motions (compared to the Sun) and 
high [$\alpha$/Fe] ratios. More recently, Neves et al. (\cite{neves09}) identified 
five planet-host stars from the 29 stars belonging to the thick disc from the HARPS GTO planet 
search programme indicating that approximately 17\% of the thick disc stars are planet hosts. 
They found some indication that metal-poor ([Fe/H] $\le$ -0.2) planet-host stars originate
preferentially in the thick disc, as also suggested by Haywood (\cite{haywood08}). 
This indication is however based on small number statistics and was 
contested by a new spectroscopic analysis made by Gonzalez (\cite{gonzalez09}),
showing that according to the mass abundance of the refractory elements (Mg, Si, Fe), 
which is important for planet formation, thick disc and [Fe/H]-poor thin disc planetary host stars 
have similar distributions. This author recommends to use a so-called refactory index $[Ref]$ 
rather than $[Fe/H]$ to compare the statistics of planets around thin disc and thick disc stars. 
Because thick disk stars are alpha-enhanced, they might have less Fe than thin disk stars, 
but they have relatively more Mg and Si, so they are not as metal -poor in terms of their 
ability to form planets as they seem to be at first glance. For WASP-21b, this refractory index $[Ref]$ 
is -0.35.

\begin{acknowledgements}

The WASP Consortium comprises astronomers primarily from the Universities of Keele, 
Leicester, The Open University, Queen's University Belfast, the University of St Andrews, 
the Isaac Newton Group (La Palma), the Instituto de Astrofisica de Canarias (Tenerife) 
and the South African Astronomical Observatory. The SuperWASP-N camera is hosted by the Isaac 
Newton Group on La Palma with funding from the UK Science and Technology 
Facilities Council. We extend our thanks to the Director and staff of the Isaac 
Newton Group for their support of SuperWASP-N operations. 
Based in part on observations made at Observatoire de Haute Provence 
(CNRS), France, and on observations made with the Nordic Optical Telescope, 
operated on the island of La Palma jointly by Denmark, Finland, Iceland, 
Norway, and Sweden, in the Spanish Observatorio del Roque de los Muchachos 
of the Instituto de Astrofisica de Canarias. 
We wish to thank the ``Programme National de Plan\'etologie'' (PNP) of CNRS/INSU, the 
Swiss National Science Foundation, and the French National Research Agency (ANR-08-JCJC-0102-01) 
for their continuous support to our planet-search programs. 
FB would like to acknowledges PLS for continuous support and advice.

\end{acknowledgements}


\begin{thebibliography}{}





\bibitem[1996]{baranne96}
Baranne, A., Queloz, D., Mayor, M., et al., 1996, A\&AS, 119, 373


\bibitem[2010]{boisse10}
Boisse, I., Bouchy, F., Chazelas, B., et al., 2010, in 
New technologies for probing the diversity of brown dwarfs and 
exoplanets, Shanghai, 2009, EPJ Web of Conferences, in press 


\bibitem[2010]{borde10}
Bord\'e, P., Bouchy, F., Deleuil, M., et al., 2010, A\&A, in press

\bibitem[2004]{bouchy04}
Bouchy, F., Pont, F., Santos, N.C., et al., 2004, A\&A, 421, L13

\bibitem[2009]{bouchy09}
Bouchy, F., H\'ebrard, G., Udry, S., et al., 2009, A\&A, 505, 853



\bibitem[2006]{cameron06}
Collier Cameron, A., Pollacco, D., Street, R.A., et al., 2006, MNRAS, 373, 799

\bibitem[1993]{edvardsson93}
Edvardsson, B., Andersen, J., Gustafsson, B., et al., 1993, A\&A, 275, 101


\bibitem[2000]{girardi00} 
Girardi, L., Bressan, A., Bertelli, G., \& Chiosi, C.\ 2000, A\&AS, 141, 371

\bibitem[1988]{gray88} 
Gray, D.F., 1988, Lectures on Spectral-line Analysis: F, G, and K Stars 
(Arva, Ontario: Publisher)

\bibitem[2006]{guillot06}
Guillot, T., Santos, N.C, Pont, F., et al., 2006, A\&A, 453, L21

\bibitem[2009]{gonzalez09}
Gonzalez, G., 2009, MNRAS, 399, L103

\bibitem[2009]{hartman09}
Hartman, J., Bakos, G., Torres, G., et al., 2009, \apj, 706, 785

\bibitem[2008]{haywood08}
Haywood, M., 2008, A\&A, 482, 673

\bibitem[2009]{hebb09} 
Hebb, L., Collier-Cameron, A., Triaud, A.H.M.J., et al.\ 2009, \apj, 693, 1920

\bibitem[2008]{hellier08}
Hellier, C., Anderson, D.R., Gillon, M., et al., 2008, \apj, 690, L89

\bibitem[2010]{hellier10}
Hellier, C., Anderson, D.R., Collier Cameron, A., et al., 2010, \apj, in press

\bibitem[2003]{mayor03}
Mayor, M., Pepe, F., Queloz, D., et al., 2003, Messenger, 114, 20

\bibitem[1990]{MFF90} 
Morrison, H.~L., Flynn, C., \& Freeman, K.~C.\ 1990, \aj, 100, 1191 

\bibitem[1997]{munari97}
Munari, U., \& Zwitter, T., 1997, A\&A, 318, 269

\bibitem[2009]{neves09}
Neves, V., Santos, N.C., Sousa, S.G., et al., 2009, A\&A, 497, 563

\bibitem[2002]{pepe02}
Pepe, F., Mayor, M. Galland, F., et al., 2002, A\&A, 388, 632

\bibitem[2008]{perruchot08}
Perruchot, S., Kohler, D., Bouchy, F., et al., 2008, in \textit{Ground-based and Airborn
Instrumentation for Astronomy II}, Edited by McLean, I.S., Casali, M.M., Proceedings of the 
SPIE, vol. 7014, 70140J

\bibitem[2006]{pollacco06}
Pollacco, D.L., Skillen, I., Cameron, A.C., et al., 2006 ,PASP, 118, 1407

\bibitem[2008]{pollacco08}
Pollacco, D., Skillen, I., Collier Cameron, A., et al., 2008, MNRAS, 385, 1576

\bibitem[2000]{queloz00}
Queloz, D., Mayor, M., Webber, L., et al., 2000, A\&A, 354, 99

\bibitem[2007]{reid07}
Reid, I.N., Turner, E.L., Turnbull, M.C., et al., 2007, \apj, 665, 767

\bibitem[2003]{besancon} 
Robin, A.~C., Reyl{\'e}, C., Derri{\`e}re, S., \& Picaud, S.\ 2003, \aap, 409, 523

\bibitem[2005]{sato05}
Sato, B., Fischer, D., Henry, G., et al., 2005, \apj, 633, 465 

\bibitem[2007]{stempels07}
Stempels, H.C., Collier Cameron, A., Hebb L., et al., 2007, MNRAS, 379, 773

\bibitem[2010]{torres10}
Torres, G., Andersen, J. \& Gimenez, A., 2010, A\&ARv, 18, 67

\bibitem[1996]{valenti96}
Valenti, J.A., \& Piskunov, N., 1996, A\&AS, 118, 595

\bibitem[2005]{valenti05}
Valenti, J.A., \& Fisher, D., 2005, ApJS, 159, 141

\bibitem[2004]{NOMAD} 
Zacharias, N., Monet, D.~G., Levine, S.~E., et al., 2004, Bulletin of the 
American Astronomical Society, 36, 1418


\end{thebibliography}
\end{document}